\documentclass[%
 reprint,
showpacs,preprintnumbers,
 amsmath,amssymb,
 aps,
prl,
]{revtex4-1}
\usepackage{graphicx}
\usepackage{dcolumn}

\begin{document}

\title{Anomalous Hall effect in two-phase semiconductor structures: the crucial role of ferromagnetic inclusions}
\date{\today}
\author{A.V. Kudrin}
\email{kudrin@nifti.unn.ru}
\affiliation{Department of Physics,
University of Nizhny Novgorod, Nizhny Novgorod 603950, Russia}
\author{A.V. Shvetsov}
\affiliation{Department of Physics, University of Nizhny Novgorod,
Nizhny Novgorod 603950, Russia}
\author{Yu.A. Danilov}
\affiliation{Department of Physics, University of Nizhny Novgorod,
Nizhny Novgorod 603950, Russia}
\author{A.A. Timopheev}
\affiliation{Department of Physics and I3N, University of Aveiro,
Aveiro 3810-193, Portugal}
\author{D.A. Pavlov}
\author{A.I. Bobrov}
\affiliation{Department of Physics, University of Nizhny Novgorod,
Nizhny Novgorod 603950, Russia}
\author{N.V. Malekhonova}
\affiliation{Department of Physics, University of Nizhny Novgorod,
Nizhny Novgorod 603950, Russia}
\author{N.A. Sobolev}
\affiliation{Department of Physics and I3N, University of Aveiro,
Aveiro 3810-193, Portugal}

\begin{abstract}
The Hall effect in InMnAs layers with MnAs inclusions of $20$-$50$
nm in size is studied both theoretically and experimentally. We
find that the anomalous Hall effect can be explained by the
Lorentz force caused by the magnetic field of ferromagnetic
inclusions and by an inhomogeneous distribution of the current
density in the layer. The hysteretic dependence of the average
magnetization of ferromagnetic inclusions on an external magnetic
field results in a hysteretic dependence of
$R_{\text{H}}(H_{ext})$. Thus we show the possibility of a
hysteretic $R_{\text{H}}(H_{ext})$ dependence (i.e. observation of
the anomalous Hall effect) in thin conductive layers with
ferromagnetic inclusions in the absence of carriers spin
polarization.
\end{abstract}
\pacs{61.72.uj, 72.20.My, 75.50.Pp} \maketitle

The investigation of the anomalous Hall effect (AHE) is a widely
used experimental method for the diagnostics of the magnetic and
transport properties of ferromagnetic layers, in particular, those
of diluted magnetic semiconductors (DMS)~\cite{Hsu}. In the
conventional interpretation, the AHE is a consequence of an
asymmetric scattering of spin-polarized charge carriers in
ferromagnetic materials~\cite{Nag}. Thus the observation of the
AHE is traditionally considered to be a proof of the presence of
spin-polarized carriers. The spin polarization of carriers in DMS
is usually attributed to a mechanism of indirect exchange
interaction between transition metal ions via charge
carriers~\cite{Dietl,Mac}. In the high temperature region this
mechanism should become slack~\cite{Dietl,Mac}. However, the AHE
was observed at room temperature or above in some Mn-doped
semiconductors~\cite{Chen,Par,Jia}. The AHE was also observed at
about 300 K in Co-doped TiO$_2$~\cite{Shin,Lee} and
(La,Sr)TiO$_3$~\cite{Zhang} layers containing Co clusters. In
Refs.~\cite{Shin,Lee,Zhang} the appearance of AHE was related to
spin polarization by extrinsic (induced by the clusters) spin
orbit scattering. Earlier, it was also observed that in InMnAs
layers obtained by laser deposition in gas atmosphere a clear
hysteresis in the magnetic field dependencies of the Hall
resistance manifests itself up to room temperature~\cite{Dan}.

In III-Mn-V layers the second-phase inclusions may appear during
technology processes. In particular, nanosize ferromagnetic MnAs
particles can be embedded in a semiconductor matrix
~\cite{Boe,Well}. The temperature dependences of the ferromagnetic
resonance~\cite{Dan} and magnetization~\cite{Dan2} for the InMnAs
layers grown by laser deposition show a Curie temperature of about
330 K that is close to the Curie temperature for MnAs. It was
shown in Ref.~\cite{Dan} that an increase of the Mn content,
characterized by the technological parameter
$Y_{\text{Mn}}$=$t_{\text{Mn}}/(t_{\text{Mn}}+t_{\text{InAs}})$,
where $t_{\text{Mn}}$ and $t_{\text{InAs}}$ are the ablation times
of the Mn and InAs targets, leads to a reinforcement of the
hysteretic character of the $R_{\text{H}}(H)$ dependences.

\begin{figure}[b]
\includegraphics{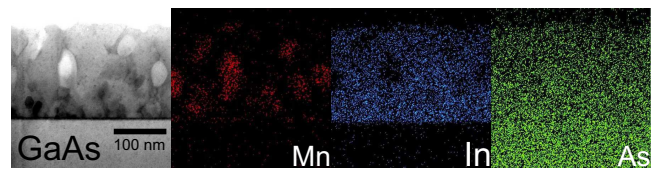}
\caption{\label{fig1} (color online) The bright-field STEM image
of the InMnAs/GaAs structure (left). The corresponding EDS
mappings for Mn, In and As.}
\end{figure}

In this Letter, we present the results of theoretical and
experimental investigations of the AHE in the InMnAs layers
obtained by laser deposition. Nevertheless our results can be
generalized to other conductive layers with ferromagnetic
inclusions. Figure~\ref{fig1} shows the bright-field
cross-sectional scanning transmission electron microscopy (STEM)
image of the InMnAs/GaAs structure with $Y_{\text{Mn}}$=0.2. The
image reveals a phase inhomogeneity of the InMnAs layer.
Figure~\ref{fig1} also shows the energy-dispersive X-ray
spectroscopy (EDS) mapping of Mn, In and As in the structure. The
bright areas in the Mn mapping image correspond to the regions of
predominantly Mn atoms. At the same time these regions are free
from In atoms. Taking into account the uniform distribution of As
atoms it can be concluded that the bright areas at Mn mapping
image correspond to the inclusions of a MnAs phase. Thus the
InMnAs layers contain MnAs clusters of 20-50 nm size.

Let us assume that the semiconductor matrix has no magnetic
ordering at the temperatures considered (200-300 K) and the charge
carriers have no predominant spin polarization. This assumption is
based on the theory of carrier-mediated ferromagnetism which
demands much higher carrier concentrations for the appearance of a
ferromagnetic ordering in DMS than those found in our InMnAs
layers ($6\cdot 10^{15} - 5\cdot 10^{18}$ cm$^{-3}$ as will be
shown below). In this case the magnetic properties of this
structure are determined only by the ensemble of ferromagnetic
MnAs inclusions. We also neglect the possible presence of any
spin-dependent processes, even if the magnetic nanoparticles
participate in the transport of electric charges.

The Lorentz force arises from the interaction of the magnetic
fields of the ferromagnetic nanoparticles and the movements of the
charge carriers. Moreover, an inhomogeneous distribution of the
current density should be observed due to the layer heterogeneity.
The principal idea of this Letter is to show that such a simple
model is sufficient to describe qualitatively and quantitatively
the hysteretic dependences of the Hall resistance on an external
magnetic field in a conductive non-magnetic matrix containing
ferromagnetic inclusions, without taking into consideration any
spin-related phenomena.

The magnetic properties of an ensemble of ferromagnetic particles
are dependent mainly on the magnetic anisotropy of the particle
$K$ and the particle volume $V_p$. These parameters define a
blocking temperature $T_{\text{B}}$ of the ensemble which
separates the superparamagnetic state from the blocked one
according to the relationship ~\cite{Neel}
$T_{\text{B}}=KV_p/25k_{\text{B}}$, where $k_{\text{B}}$ is the
Boltzmann constant. As is known~\cite{Blois}, MnAs has an
"easy-plane" magnetocrystalline anisotropy for which
$T_{\text{B}}$ should be 0 K. Thus it cannot provide the blocked
character of the magnetization reversal with the coercivity
$H_{\text{C}}\sim 0.5$ kOe, as it was observed for our InMnAs
layers~\cite{Dan,Dan2} as well as for 20-100 nm size MnAs clusters
embedded in a GaAs matrix~\cite{Boe,Well}. At the same time, the
particles have an elongated shape (see Fig.~\ref{fig1}) and in our
case it is apparently the main source of the magnetic hysteresis.
The microscopy revealed that the typical particle shape is an
ellipsoid with the axes $a\simeq b \simeq c/2$. This results in
the effective uniaxial anisotropy constant $K\simeq 5\cdot 10^5$
erg/cm$^3$ for $M_{\text{S}}=$600 emu/cm$^3$ (Ref.~\cite{Blois}).
So, to be unblocked at room temperature, the particle size must be
less than 15 nm. The observed particles are several times larger
than this (Fig.~\ref{fig1}). Therefore the considered ensemble of
the MnAs nanoparticles is in the blocked state at room temperature
and thus possesses a coercivity.
\begin{figure}[t]
\includegraphics{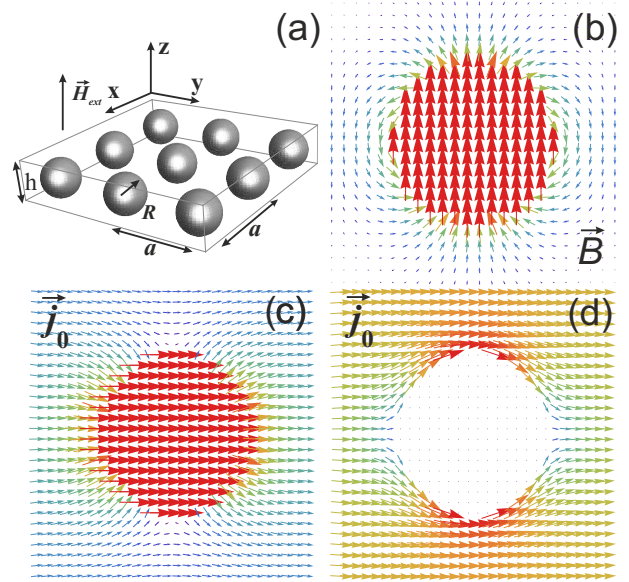}
\caption{\label{fig2} (color online) (a) The two-dimensional
lattice of spherical ferromagnetic particles ($a$ is the
interparticle distance, $R$ is the radius of the particles); (b)
The distribution of the magnetic induction $\vec B$ in the layer.
(c) The current lines for the case of particles with high
conductivity. (d) The current lines for the case of particles with
low conductivity.}
\end{figure}

In the $z$ direction (normal to the layer plane), the film
contains one or a few MnAs particles (Fig.~\ref{fig1}). Thus a
two-dimensional lattice of spherical ferromagnetic particles can
be taken as the simplest model of the considered system
(Fig.~\ref{fig2}(a)). If the system is classically described and
there is no barrier for carriers at the particle/matrix interface
then the semi-metallic MnAs particles in a semiconductor matrix
can be considered as a region with enhanced conductivity. We
assume that a current flows in the $y$ direction then the current
density outside the particle is
\begin{equation}\label{eq1}
    \vec j=\vec j_0
    \left(1-\frac{R^3}{r^3}\frac{\sigma_1-\sigma_2}{\sigma_1+2\sigma_2}\right)+
    \frac{\sigma_1-\sigma_2}{\sigma_1+2\sigma_2}\frac{3j_0R^3y\vec
    r}{r^5},
\end{equation}
while the current density inside the particle is
\begin{equation}\label{eq2}
    \vec j=\frac{3\sigma_1}{\sigma_1+2\sigma_2}\vec j_0,
\end{equation}
where $\vec j_0$ denotes the current density far from a particle.
The origin of coordinates is in the center of a particle. The
conductivity of the semiconductor matrix is characterized by a
value $\sigma_2$ and the conductivity of the particle is
$\sigma_1$. Since $\sigma_1>\sigma_2$, the current lines are
involved into the particle as shown in Fig.~\ref{fig2}(c) and the
current density in the particle is distinctly higher than far from
it. On the other hand the particle/matrix interface can be a
barrier for carriers. Consequently the particle can be considered
as a region with low effective conductivity ($\sigma_1<\sigma_2$)
and the current flow will bend around the inclusion
(Fig.~\ref{fig2}(d)). In the limiting case when the barrier is
impenetrable, charge carriers move in a space between the
inclusions in a magnetic field produced by the inclusions and in
an external magnetic field. It will be shown hereafter that,
regardless of whether there is a barrier or it is absent, the
hysteretic dependence of the Hall resistance on an external
magnetic field ($R_{\text{H}}(H_{ext})$) will be observed. This is
attributed to the system heterogeneity and to the hysteretic
dependence of the inclusion magnetization on an external magnetic
field.

For the layer in the external magnetic field $\vec H_{ext}$
applied along the $z$ direction the average force acting on
carriers per unit volume is defined as
\begin{equation}\label{eq4}
    \vec F=\frac{1}{\Delta V}\int_{\Delta V}\frac{1}{c}\left[\vec j \times \vec B\right] dV,
\end{equation}
where $\Delta V=ha^2$ is the volume of the layer per ferromagnetic
particle. Assuming that all the particles are uniformly magnetized
along the $z$ direction each particle produces the magnetic field
at the point $\vec r_i$ equal to
\begin{equation}\label{eq5}
    \vec H_i=\frac{3\left(\vec m \vec r_i\right)\vec
    r_i}{r_i^5}-\frac{\vec m}{r_i^3},
\end{equation}
where $\vec m=(4/3)\pi R^3\vec M_0$ is the magnetic moment of the
particle, $\vec M_0$ is the magnetization of the latter. The
particle with the number $j$ is subject to the external magnetic
field $\vec H_{ext}$ and the fields of the other
particles~(\ref{eq5}). Assuming that the distance $a$ between the
particles noticeably exceeds $R$ the field produced by all other
particles near the $j$-th particle can be considered as locally
homogeneous $\vec H_p\simeq -(2\pi h/a)\vec M_0\vartheta$, where
$\vartheta$ is the volume fraction of the ferromagnetic
inclusions. Consequently, the field inside each particle is
\begin{equation}\label{eq7}
    \vec H\simeq\vec H_{ext}+\vec H_p-(4/3)\pi\vec M_0.
\end{equation}
The field in the space between the particles cannot be considered
as homogeneous, but hereinafter we will use only the average value
of the field. Since the volume per particle is $ha^2$ and all the
particles are identical, we find the average field $\bar{H}_z$ in
the space between the particles using the superposition principle
\begin{equation}\label{eq8}
    \bar{H}_z=H_{ext}+\frac{1}{ha^2}\int_V H_{i,z}dV,
\end{equation}
where integration is over the layer volume. The substitution of
$\vec H_i$  Eq.~(\ref{eq5}) into Eq.~(\ref{eq8}) yields
\begin{equation}\label{eq9}
    \bar{H}_z=H_{ext}-\frac{8\pi}{3}M_0\vartheta,
\end{equation}
and $\bar H_x=\bar H_y=0$ owing to the problem symmetry.

\begin{figure}[t]
\includegraphics{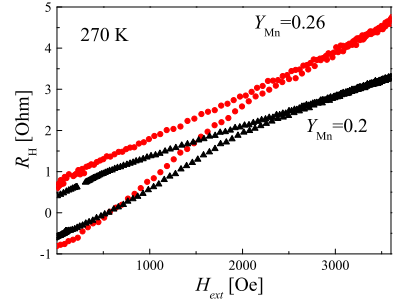}
\caption{\label{fig3}(color online) The $R_{\text{H}}(H_{ext})$
dependences measured at 270 K for the InMnAs layer with manganese
content $Y_{\text{Mn}}$=0.2 (black triangles) and
$Y_{\text{Mn}}$=0.26 (red circles).}
\end{figure}

Let us examine two limit cases. (i) Let a barrier for carriers at
the particle/matrix interface be present, so $\sigma_1 <<
\sigma_2$. In this case as it follows from Eq.~(\ref{eq2}) $\vec
j\sim(\sigma_1/\sigma_2)\vec j_0\approx 0$ and the current flow
does not penetrate into the inclusion. Therefore the charge
carriers move in the space between the inclusions in the average
field $\bar{H}_z$ Eq.~(\ref{eq8}). (ii) If the barrier is absent
then $\sigma_1 >> \sigma_2$. Hence the current density in the
particle is $\vec j\simeq 3\vec j_0$ (Eq.~(\ref{eq2})) and the
current flow is involved into the particle. In this case carriers
inside the particle are subject to the magnetic field
Eq.~(\ref{eq7}). For the case (i) the average Lorentz force can be
approximately found in according to Eqs.~(\ref{eq4}),~(\ref{eq9})
as
\begin{equation}\label{eq10}
    F_x\simeq\frac{1}{ha^2}\int_{V_m}\frac{1}{c}j_0\bar H_z
    dV=-\frac{8\pi}{3c}j_0M_0\vartheta+\frac{1}{c}j_0H_{ext},
\end{equation}
and for the case (ii) according to
Eqs.~(\ref{eq4}),~(\ref{eq7}),~(\ref{eq9}) as
\begin{eqnarray}\label{eq11}
    F_x\simeq\frac{1}{ha^2c}\Big(\int_{V_p}j_0\left(H_{ext}+H_p+\frac{8\pi}{3}M_0\right)dV+\nonumber\\+\int_{V_m}j_0\bar H_z
    dV\Big)
    \simeq\frac{16\pi}{3c}j_0M_0\vartheta+\frac{1}{c}j_0\left(1+\vartheta\right)H_{ext},
\end{eqnarray}
where $V_p$ is the particle volume and $V_m=ha^2-V_p$. According
to Eqs.~(\ref{eq10}) and~(\ref{eq11}), the average Lorentz force
depends on the magnetization of the particles. It should be noted
that for the case (i) the sign of the first term in
Eq.~(\ref{eq10}) is different from that for the case (ii)
Eq.~(\ref{eq11}). This will be discussed in more detail hereafter.
The MnAs inclusions are ferromagnetic and the hysteretic
dependence of the magnetization $M_0$ on the external magnetic
field leads to a hysteretic dependence of the Hall resistivity on
the external magnetic field
$\rho_{\text{H}}(H_{ext})=F_x(H_{ext})/(enj_0)$, where $n$ is the
charge carrier concentration in the semiconductor matrix, and $e$
is the electron charge. According to Eqs.~(\ref{eq10})
and~(\ref{eq11}) the remanent Hall resistance is
\begin{equation}\label{eq13}
    R_{\text{H}}(H_{ext}=0)\simeq\gamma\frac{8\pi}{3}\frac{M_0\vartheta}{hnec},
\end{equation}
where $\gamma=1$ for $\sigma_1<<\sigma_2$ and $\gamma=2$ for
$\sigma_1>>\sigma_2$. As an estimation of $M_0$ we take the value
of the remanent magnetization of an epitaxial MnAs layer (about
900 G at 270 K~\cite{Xu}). The typical value of the carrier
concentration for our InMnAs layers at 270 K is about $3\cdot
10^{18}$ cm$^{-3}$ and the layer thickness $h\simeq$ 190 nm. For
the volume fraction of the MnAs inclusions in the InMnAs layer is
about 0.05 (Fig.~\ref{fig1}), the value of
$R_{\text{H}}(H_{ext}=0)$ equals approximately 0.5 Ohm, which is
in a good agreement with the experimental results
(Fig.~\ref{fig3}).

Since for the discussed model the hysteresis in the
$R_{\text{H}}(H_{ext})$ dependence is due to the Lorentz force, it
can be attributed to the ordinary Hall effect (OHE). Consequently,
for a fixed volume fraction and a magnetization value of the MnAs
particles in the layer, the value of the remanent Hall resistance
should increase with decreasing the carrier concentration, in
accordance with Eq.~(\ref{eq13}).

\begin{table}[t]
\caption{The concentration of charge carriers at 300 and 200 K in
the InMnAs layer with $Y_{\text{Mn}}$=0.2 before and after
implantation of protons with different fluences. In parentheses we
indicate the type of majority carriers.\label{table}}
\begin{ruledtabular}
\begin{tabular}{ccc}
 \multicolumn{1}{c}{Fluence (cm$^{-2}$)}&\multicolumn{1}{c}{Carrier concentration}&\multicolumn{1}{c}{Carrier concentration}\\
 &at 300 K (cm$^{-3}$)&at 200 K (cm$^{-3}$)
\\
 \hline
 0& $4.8\cdot 10^{18}$ ($p$) & $3.5\cdot 10^{18}$ ($p$)\\
  $1\cdot 10^{13}$ & $3.5\cdot 10^{18}$ ($p$) & $2.3\cdot 10^{18}$ ($p$)\\
  $3\cdot 10^{13}$ & $9.8\cdot 10^{17}$ ($p$) & $5.5\cdot 10^{17}$ ($p$)\\
  $1\cdot 10^{14}$ & $2.7\cdot 10^{16}$ ($n$) & $6.0\cdot 10^{15}$ ($n$)\\
  $2.1\cdot 10^{14}$ & $1.0\cdot 10^{17}$ ($n$)& $4.7\cdot 10^{16}$ ($n$)\\
  $6\cdot 10^{14}$ & $1.6\cdot 10^{17}$ ($n$)& $9.0\cdot 10^{16}$ ($n$)
\end{tabular}
\end{ruledtabular}
\end{table}
It is known that the ion irradiation can vary the carrier
concentration of in semiconductors due to the formation of
radiation-induced crystal defects. A feature of the InAs
semiconductor is that radiation defects shift the Fermi level
toward the conduction band, which leads to an increase of the
carrier concentration in the n-type material or the p-n type
conversion in p-InAs (Ref.~\cite{Brud}). To change the carrier
concentration in the InMnAs layer, the proton implantation was
carried out with an energy of 50 keV and a fluence in the range
$1\cdot 10^{13}$ - $6\cdot 10^{14}$ cm$^{-2}$. Table~\ref{table}
shows the values of the carrier concentration at 300 and 200 K in
the InMnAs layer with $Y_{\text{Mn}}$=0.2 before and after
irradiations with different proton fluences. The values of the
carrier concentration were determined from the slope of the
$R_{\text{H}}(H_{ext})$ dependences in $H_{ext}$ above 3000 Oe,
i.e. mainly in the linear region. The temperature of 200 K was the
lowest one at which it was possible to obtain the
$R_{\text{H}}(H_{ext})$ dependences for a high resistance layer
irradiated by protons with a fluence of $1\cdot 10^{14}$
cm$^{-2}$.

Proton implantations with fluences of $1\cdot 10^{13}$ and $3\cdot
10^{13}$ cm$^{-2}$ lead to a decrease in the concentration of
carriers (holes) as a result of the partial compensation of the Mn
acceptor impurity by radiation-induced donor-type defects
(Table~\ref{table}). The conversion of the conductivity type from
p to n is observed after the implantation with a fluence of
$1\cdot 10^{14}$ cm$^{-2}$. With a further increase in the proton
fluence to $6\cdot 10^{14}$ cm$^{-2}$ the concentration of the
majority carriers (electrons) rises (Table~\ref{table}). We
suppose that the proton implantation does not lead to a
significant modification of the magnetic properties of the
semi-metallic MnAs inclusions.

\begin{figure}[t]
\includegraphics{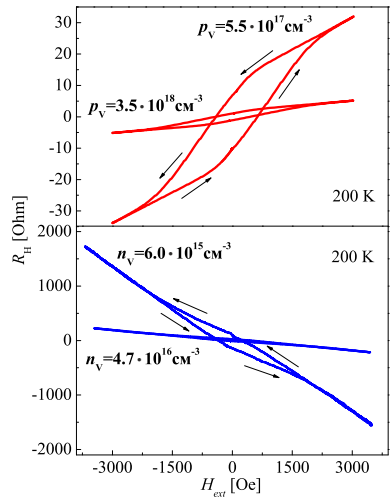}
\caption{\label{fig4}(color online) The $R_{\text{H}}(H_{ext})$
dependences measured at 200 K for the InMnAs layer
($Y_{\text{Mn}}$=0.2) for the different carrier concentration
values. The arrows indicate the magnetic field scan directions.}
\end{figure}

Figure~\ref{fig4} shows the $R_{\text{H}}(H_{ext})$ dependences at
200 K for the InMnAs layer with different carrier concentration
values. For a fixed $H_{ext}$ the Hall resistance value of the
linear part of the $R_{\text{H}}(H_{ext})$ dependence increases
with decreasing carrier concentration. This is typical of the OHE
and is related to the increase of the Hall coefficient $R_0=1/en$.
We emphasize that both for the cases of the p- and n-type majority
carriers the clear increase of the remanent Hall resistance with
decreasing carrier concentration is also observed
(Fig.~\ref{fig4}). So, the experimental results are in a good
agreement with our model: the $R_{\text{H}}(H_{ext}=0)$ value
increases with decreasing carrier concentration.
\begin{figure}
\includegraphics{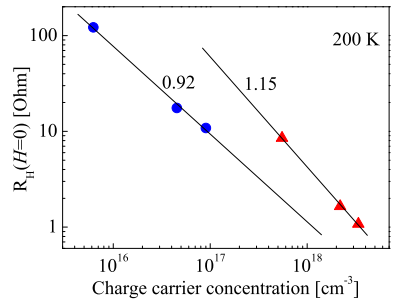}
\caption{\label{fig5}(color online) The experimental dependences
of the remanent Hall resistance on the hole concentration
(triangles) and electron concentration (circles).  The solid lines
show linear approximations of the experimental data.}
\end{figure}

Figure~\ref{fig5} shows the experimental dependences of the
remanent Hall resistance on the carrier concentration at a
temperature of 200 K. Since for the OHE the Hall resistance is
inversely proportional to the carrier concentration, the
dependence of the remanent Hall resistance on the concentration in
the double-logarithmic coordinates should be linear
(Fig.~\ref{fig5}). The linear approximation of the experimental
points is shown in Fig.~\ref{fig5}. The slope coefficients are
sufficiently close to unity (as shown in Fig.~\ref{fig5}) which
corresponds to the assumption about the determinative role of the
OHE (Eq.~\ref{eq13}) in the observed hysteretic dependences
$R_{\text{H}}(H_{ext})$ ~\cite{Ber}. We note that for a fixed
carrier concentration the remanent Hall resistance will be higher
for the p-type carriers than for the n-type carriers
(Fig.~\ref{fig5}). This effect has been predicted by the proposed
model. It follows from Eqs.~(\ref{eq10})--~(\ref{eq13}) that in
the case when the conductivity of the inclusions is higher than
that of the matrix, the influence of the magnetic field of the
inclusions on the Hall effect will be two times higher than in the
opposite case. Thus it is possible to conclude that for holes the
barrier at the cluster/matrix interface is absent. However for
electrons an impenetrable barrier at the cluster/matrix interface
is present, which causes their predominant movement in the InMnAs
layer between the MnAs clusters. It is known that MnAs is a p-type
ferromagnetic semi-metal. This is in a good agreement with our
conclusion about the presence of a barrier for electrons between
the InMnAs matrix and the MnAs cluster.

It should be noted that, for p-type carriers, the sign of the
linear component of the $R_{\text{H}}(H_{ext})$ dependencies
coincides with the sign of the hysteretic component. For n-type
carriers these components have a different sign (Fig.~\ref{fig4}).
The difference between the signs of the linear component is
related to the different type of carriers. At the same time, both
for the p- and n-type majority carriers the sign of the hysteretic
component is the same (Fig.~\ref{fig4}). The reason for that is
the following. The holes are affected by the Lorentz force. This
leads to the average Lorentz force given by Eq.~\ref{eq11}. Since
the direction of the magnetization of the MnAs clusters (at the
saturation point) coincides with that of $H_{ext}$
(Fig.~\ref{fig2}(b)) the sign of the hysteretic component of the
$R_{\text{H}}(H_{ext})$ dependence (which is determined by the
magnetization of the MnAs inclusions) is equal to the sign of the
linear component which is determined by $H_{ext}$. In contrast, in
n-type layers the majority carriers (electrons) move between the
MnAs inclusions in the magnetic field of inclusions and in the
external magnetic field. However the magnetic field which is
produced by the clusters has the opposite direction to the
magnetization of the clusters (Fig.~\ref{fig2}(b)) and is
consequently opposite to $H_{ext}$. This leads to the difference
in the signs of the terms in equation~(\ref{eq10}) for the average
Lorentz force and to the difference in the signs of the linear and
hysteretic components of the $R_{\text{H}}(H_{ext})$ dependencies
for n-type layers (Fig.~\ref{fig4}).

Based on all the issues discussed above we can conclude that the
apparent AHE in thin conductive layers with ferromagnetic
inclusions can be related to the influence of local magnetic
fields produced by the particles on charge carriers and an
inhomogeneous distribution of the current density in the layer.
Thus we have demonstrated the possibility of the pronounced
anomalous Hall effect in the absence of the spin polarization of
charge carriers.

The authors thank B.N. Zvonkov for the growth of the samples and
V.K. Vasiliev for the proton implantation. We acknowledge support
from RFBR (grants No. 12-07-00433-a, No. 12-07-31161), FCT through
grants and projects PEst-C/CTM/LA0025/2011,
RECI/FIS-NAN/0183/2012, SFRH/BPD/74086/2010 and by the European
project "Mold-Nanonet".

\end{document}